\documentclass[prd,twocolumn,showpacs,preprintnumbers,amsmath,amssymb,superscriptaddress,floatfix,nofootinbib]{revtex4}

\usepackage{graphicx}
\usepackage{amsmath}
\usepackage{amsfonts}
\usepackage{amssymb}
\usepackage{color}
\usepackage{multirow}
\usepackage[colorlinks, citecolor=blue,anchorcolor=red,menucolor=red, linkcolor=red,filecolor=red,runcolor=red,urlcolor=blue,frenchlinks=red]{hyperref}

\begin{document}

\title{Excited bottom and bottom-strange mesons in the quark model}

\author{Qi-Fang L\"{u}, Ting-Ting Pan, Yan-Yan Wang, En Wang, and De-Min Li} \email{lidm@zzu.edu.cn}
\affiliation{Department of Physics, Zhengzhou University, Zhengzhou, Henan 450001, China}

\begin{abstract}

 In order to understand the possible $q\bar{q}$ quark-model assignments of the $B_J(5840)$ and $B_J(5960)$ recently reported by the LHCb Collaboration, we evaluate mass spectra, strong decays, and radiative decays of bottom and bottom-strange mesons in a nonrelativistic quark model. Comparing these predictions with the relevant experimental results, we suggest that the $B_J(5840)$ and $B_J(5960)$ can be identified as $B(2^1S_0)$ and $B(1^3D_3)$, respectively, and the $B(5970)$ reported by the CDF Collaboration can be interpreted as $B(2^3S_1)$ or $B(1^3D_3)$. Further precise measurements of the width, spin and decay modes of the $B(5970)$ are needed to distinguish these two assignments. These predictions of bottom and bottom-strange mesons can provide useful information to further experimental investigations.
\end{abstract}
 \maketitle

\section{Introduction}{\label{introduction}}

Heavy-light mesons composed of one heavy quark and one light quark act as the hydrogen atoms of hadron physics and are the ideal laboratory for the understanding of strong interactions in the non-perturbative regime\cite{Derujula76,rosner86,godfreyRMP}.
In the past several years, significant progress has been achieved in studying the charmed and charmed-strange states exerimentally\cite{Agashe:2014kda,Aaij:2014xza,Aaij:2014baa,Aaij:2015vea,Aaij:2015sqa}. It is widely accepted that the ground charmed and charmed-strange mesons such as $D(1S)$, $D(1P)$, $D_s(1S)$, and $D_s(1P)$ have been established\cite{Agashe:2014kda}, and some candidates for higher radial and orbital excitations have also been reported, which have stimulated many theoretical
   investigations on these excitations\cite{Zhong:2010vq,zhong2,Xiao:2014ura,Lu:2014zua,Wang:2010ydc,fazio2012,Wang:2013tka,Wang:2014jua,Sun:2010pg,Li:2010vx,Sun:2013qca,Song:2014mha,Song:2015nia,Close:2005se,Godfrey:2014fga,Segovia:2012cd,Segovia:2015dia,othercharm1,othercharm2,othercharm3}.

Recently, the LHCb Collaboration studied $B^+\pi^-$ and $B^0\pi^-$ invariant mass distributions by analysing $pp$ collision data at centre-of-mass energies of 7 and 8 TeV\cite{Aaij:2015qla}. Precise masses and widths of the $B_1(5721)$ and $B_2^*(5747)$ are measured, and two excited bottom mesons $B_J(5840)^{0,+}$ and $B_J(5960)^{0,+}$ are observed, whose masses and widths are also studied with various quantum number hypotheses. The measured masses and widths of neutral $B_J(5840)$ and $B_J(5960)$ under different spin-parity hypotheses are listed in Table \ref{exp}. In 2013, the CDF Collaboration studied the $B^0\pi^+$ and $B^+\pi^-$ invariant mass distributions using the data from $p\bar{p}$ collisions at $\sqrt{s} = 1.96$ TeV\cite{Aaltonen:2013atp}. A new resonance $B(5970)$ are found both in the $B^0\pi^+$ and $B^+\pi^-$ mass distributions, whose mass and width of the neutral state are $5978\pm 5\pm 12$ MeV and $70^{+30}_{-20}\pm30$ MeV, respectively. Since the $B(5970)$ can decay into $B\pi$ final state, it should be a natural spin-parity state.

Unlike the prosperity of charm sector, experimental information on excited bottom and bottom-strange mesons is scarce. Therefore, the above excited $B$ mesons reported by the LHCb and CDF Collaborations provide a good platform to study the low-lying excited bottom and bottom-strange mesons. Some theoretical predictions on masses and widths of bottom and bottom-strange mesons have been performed in different approaches such as constituent quark model\cite{Godfrey:1985xj,zvr,de,efg,Lahde:1999ih}, chiral quark model\cite{Xiao:2014ura}, $^3P_0$ model\cite{Sun:2014wea,Luo:2009wu,Ferretti:2015rsa}, heavy meson effective theory\cite{Xu:2014mka,Wang:2014cta} and other approaches\cite{Gelhausen:2014jea,Wang:2014oca}. These theoretical predictions are not completely consistent with each other. In order to under the natures of the $B_J(5840)$, $B_J(5960)$, and $B(5970)$, further test calculations against
the experimental measurements are required.

The main purpose of this work is to discuss the possible quark-model assignments of the $B_J(5840)$, $B_J(5960)$, and $B(5970)$. We shall calculate the masses of excited bottom and bottom-strange mesons in a nonrelativistic quark model and the corresponding strong decay behaviors in the $^3P_0$ model. The relevant radiative transitions are also evaluated.

This work is organized as follows. In Sec. II, we calculate the bottom and bottom-strange meson masses in a nonrelativistic quark model. In Sec. III, we evaluate the two-body OZI allowed strong decays of the bottom and bottom-strange mesons in the $^3P_0$ model with the realistic wave functions from the quark model employed in Sec. II.  In Sec. IV, we give the $E1$ and $M1$ radiative decays of the bottom and bottom-strange mesons. A summary is given in the last section.

\begin{table}[htb]
\begin{center}
\caption{ \label{exp} The neutral charge resonances observed by the
LHCb Collaboration with different spin-parity hypotheses\cite{Aaij:2015qla}. The N and UN stand
for the natural spin-parity $[P=(-1)^J]$ and unnatural spin parity $[P=(-1)^{(J+1)}]$, respectively. }
\footnotesize
\begin{tabular}{lc}
\hline\hline
 \multicolumn{2}{c}{hypothesis I: Both $B_J(5840)^0$ and $B_J(5960)^0$ have UN.  }\\\hline
 $M_{B_J(5840)^0}$     & $5862.9\pm 5.0\pm 6.7\pm0.2$ MeV        \\
 $\Gamma_{B_J(5840)^0}$ &  $127.4\pm 16.7 \pm 34.2$MeV\\
 $M_{B_J(5960)^0}$     & $5969.2\pm 2.9\pm 5.1\pm0.2$ MeV\\
 $\Gamma_{B_J(5960)^0}$ &  $82.3\pm 7.7 \pm 9.4$   MeV    \\\hline
 \multicolumn{2}{c}{hypothesis II: $B_J(5840)^0$ has N and $B_J(5960)^0$ has UN. }\\\hline

 $M_{B_J(5840)^0}$     & $5889.7\pm 22.0\pm 6.7\pm0.2$MeV       \\
 $\Gamma_{B_J(5840)^0}$ &  $107.0\pm 19.6 \pm 34.2$ MeV    \\
 $M_{B_J(5960)^0}$     &  $6015.9\pm 3.7\pm 5.1\pm0.2\pm0.4$ \\
 $\Gamma_{B_J(5960)^0}$ &  $81.6\pm 9.9 \pm 9.4$  MeV            \\\hline
 \multicolumn{2}{c}{hypothesis III: $B_J(5840)^0$ has UN  and $B_J(5960)^0$ has N. }\\\hline

 $M_{B_J(5840)^0}$    & $5907.8\pm 4.7\pm 6.7\pm0.2\pm0.4$ MeV     \\
 $\Gamma_{B_J(5840)^0}$& $119.4\pm 17.2 \pm 34.2$ MeV  \\

 $M_{B_J(5960)^0}$    & $5993.6\pm 6.4\pm 5.1\pm0.2$ MeV\\
 $\Gamma_{B_J(5960)^0}$&   $55.9\pm 6.6 \pm 9.4$ MeV         \\\hline\hline
\end{tabular}
\end{center}
\end{table}

\section{Masses}{\label{masses}}

To obtain the bottom and bottom-strange meson spectroscopy, we calculate their masses
in a nonrelativistic quark model
proposed by Lakhina and Swanson, which can describe the heavy-light meson and heavy quarkonium masses with reasonable accuracy\cite{Lakhina:2006fy}. We have employed this model to evaluate the open-charm mesons masses in Ref.\cite{Li:2010vx}. In this model,
the total Hamiltonian can be written as
\begin{equation}
H = H_0+H_{sd}+C_{q\bar{q}},\label{Ham}
\end{equation}
where $H_0$ is the zeroth-order Hamiltonian, $H_{sd}$ is the spin-dependent Hamiltonian, and $C_{q\bar{q}}$ is a constant. The $H_0$ is
\begin{equation}
H_0 = \frac{\boldsymbol{p}^2}{M_r}-\frac{4}{3}\frac{{\alpha}_s}{r}+br+\frac{32{\alpha}_s{\sigma}^3 e^{-{\sigma}^2r^2}}{9\sqrt{\pi}m_qm_{\bar{q}}} {\boldsymbol{S}}_{q} \cdot {\boldsymbol{S}}_{\bar{q}},
\end{equation}
where $\boldsymbol{p}$ is the center-of-mass momentum, $r$ is the $q\bar{q}$ separation, $M_r=2m_qm_{\bar{q}}/(m_q+m_{\bar{q}})$;  $m_q$ and $m_{\bar{q}}$ are the masses of quark $q$ and antiquark $\bar{q}$, respectively; ${\boldsymbol{S}}_{q}$ and ${\boldsymbol{S}}_{\bar{q}}$ are the spins of the quark $q$ and antiquark $\bar{q}$, respectively. The spin-dependent part $H_{sd}$ can be expressed as
\begin{eqnarray}
H_{sd} &=& \left(\frac{{\boldsymbol{S}}_{q}}{2m_q^2}+\frac{{\boldsymbol{S}}_{\bar{q}}}{2m_{\bar{q}}^2}\right) \cdot \boldsymbol{L}\left(\frac{1}{r}\frac{dV_c}{dr}+\frac{2}{r}\frac{dV_1}{dr}\right)\nonumber \\
&& +\frac{{\boldsymbol{S}}_+ \cdot \boldsymbol{L}}{m_qm_{\bar{q}}}\left(\frac{1}{r} \frac{dV_2}{r}\right) \nonumber \\
&& +\frac{3{\boldsymbol{S}}_{q} \cdot \hat{\boldsymbol{r}}{\boldsymbol{S}}_{\bar{q}} \cdot \hat{\boldsymbol{r}}-{\boldsymbol{S}}_{q} \cdot {\boldsymbol{S}}_{\bar{q}}}{3m_qm_{\bar{q}}}V_3 \nonumber \\
&& +\left[\left(\frac{{\boldsymbol{S}}_{q}}{m_q^2}-\frac{{\boldsymbol{S}}_{\bar{q}}}{m_{\bar{q}}^2}\right)+\frac{{\boldsymbol{S}}_-}{m_qm_{\bar{q}}}\right] \cdot \boldsymbol{L} V_4,
\end{eqnarray}
where $\boldsymbol{L}$ is the relative orbital angular momentum of the $q\bar{q}$ system, and
\begin{eqnarray}
V_c &=& -\frac{4}{3}\frac{{\alpha}_s}{r}+br, \nonumber\\
V_1 &=& -br-\frac{2}{9\pi}\frac{{\alpha}_s^2}{r}[9{\rm ln}(\sqrt{m_qm_{\bar{q}}}r)+9{\gamma}_E-4],\nonumber\\
V_2 &=& -\frac{4}{3}\frac{{\alpha}_s}{r}-\frac{1}{9\pi}\frac{{\alpha}_s^2}{r}[-18{\rm ln}(\sqrt{m_qm_{\bar{q}}}r)\nonumber\\&&+54{\rm ln}(\mu r)+36{\gamma}_E+29],\\
V_3 &=& -\frac{4{\alpha}_s}{r^3}-\frac{1}{3\pi}\frac{{\alpha}_s^2}{r^3}[-36{\rm ln}(\sqrt{m_qm_{\bar{q}}}r)\nonumber\\&&+54{\rm ln}(\mu r)+18{\gamma}_E+31],\nonumber\\
V_4 &=& \frac{1}{\pi}\frac{{\alpha}_s^2}{r^3}{\rm ln}\left(\frac{m_{\bar{q}}}{m_q}\right),\nonumber\\
{\boldsymbol{S}}_{\pm} &=& {\boldsymbol{S}}_q\pm{\boldsymbol{S}}_{\bar{q}}\nonumber.
\end{eqnarray}
Here $\gamma_E$ = 0.5772 and the scale $\mu$ is set to 1.1 GeV.

The parameters used in this work are ${\alpha}_s = 0.5$, $b = 0.14~ {\rm GeV^2}$, $\sigma = 1.17 ~{\rm GeV}$, $C_{d\bar{b}} = 0.003~ {\rm GeV}$, $C_{s\bar{b}} = 0.051~ {\rm GeV}$. The constituent quark masses are taken to be $m_u = m_d = 0.45 ~{\rm GeV}$, $m_s = 0.55 ~{\rm GeV}$, and $m_b = 4.5 ~{\rm GeV}$.

The spin-orbit term included in the $H_{sd}$ can be decomposed into
symmetric part $H_{sym}$ and antisymmetric part $H_{anti}$. These two parts can be written as
\begin{eqnarray}
H_{sym} &=& \frac{{\boldsymbol{S}}_+ \cdot {\boldsymbol{L}}}{2}\left[\left(\frac{1}{2m_q^2}+\frac{1}{2m_{\bar{q}}^2}\right) \left(\frac{1}{r}\frac{dV_c}{dr}+\frac{2}{r}\frac{dV_1}{dr}\right)\right. \nonumber \\
&& \left.+\frac{2}{m_qm_{\bar{q}}}\left(\frac{1}{r} \frac{dV_2}{r}\right)+\left(\frac{1}{m_q^2}-\frac{1}{m_{\bar{q}}^2}\right)V_4\right],
\end{eqnarray}
\begin{eqnarray}
H_{anti} &=& \frac{{\boldsymbol{S}}_- \cdot {\boldsymbol{L}}}{2}\left[\left(\frac{1}{2m_q^2}-\frac{1}{2m_{\bar{q}}^2}\right) \left(\frac{1}{r}\frac{dV_c}{dr}+\frac{2}{r}\frac{dV_1}{dr}\right)\right. \nonumber \\
&& \left.+\left(\frac{1}{m_q^2}+\frac{1}{m_{\bar{q}}^2}+\frac{2}{m_qm_{\bar{q}}}\right)V_4\right].
\end{eqnarray}
The antisymmetric part $H_{anti}$ gives rise to the
the spin-orbit mixing of the heavy-light mesons with different total spins but with the same total angular momentum such as
$B(n{}^3L_L)$ and $B(n{}^1L_L)$ [$B_s(n{}^3L_L)$ and $B_s(n{}^1L_L)$].  Hence, the two physical states $B_L(nL)$ and $B^\prime_L(nL)$ [$B_{sL}(nL)$ and $B^\prime_{sL}(nL)$] can be
expressed as\cite{Lu:2014zua,Godfrey:1985xj,Godfrey:1986wj}
\begin{equation}
\left(
\begin{array}{cr}
B_L(nL)\\
B^\prime_L(nL)
\end{array}
\right)
 =\left(
 \begin{array}{cr}
\cos \theta_{nL} & \sin \theta_{nL} \\
-\sin \theta_{nL} & \cos \theta_{nL}
\end{array}
\right)
\left(\begin{array}{cr}
B(n^1L_L)\\
B(n^3L_L)
\end{array}
\right),
\label{Bmixing1}
\end{equation}
\begin{equation}
\left(
\begin{array}{cr}
B_{sL}(nL)\\
B^\prime_{sL}(nL)
\end{array}
\right)
 =\left(
 \begin{array}{cr}
\cos \theta^\prime_{nL} & \sin \theta^\prime_{nL} \\
-\sin \theta^\prime_{nL} & \cos \theta^\prime_{nL}
\end{array}
\right)
\left(\begin{array}{cr}
B_s(n^1L_L)\\
B_s(n^3L_L)
\end{array}
\right),
\label{Bmixing1}
\end{equation}
where the $\theta_{nL}$ and $\theta^\prime_{nL}$ are the mixing
angles. The $B^\prime_L(nL)$ [$B^\prime_{sL}(nL)$] refers to the
higher mass state.

With the help of Mathematica program\cite{Lucha:1998xc}, we solve
the Schr\"{o}dinger equation with Hamiltonian $H_0$ and treat the
$H_{sd}$ as the perturbative term. The obtained bottom and
bottom-strange meson masses are shown in Table \ref{bmass} and
\ref{bsmass}. The predictions of some other quark
models\cite{zvr,de,efg,Lahde:1999ih} are also listed.

\begin{table}[htbp]
\begin{center}
\caption{ \label{bmass} The $B$ meson masses in MeV from
different quark models. The mixing angles of $B_L-B^\prime_L$ obtained in
this work are $\theta_{1P}=-34.6^\circ$, $\theta_{2P}=-36.1^\circ$, $\theta_{1D}=-39.6^\circ$,
$\theta_{2D}=-39.7^\circ$, $\theta_{1F}=-41.0^\circ$. A dash denotes
that the corresponding mass was not calculated in the corresponding reference.}
\footnotesize
\begin{tabular}{lccccc}
\hline\hline
  State         & This work  & ZVR\cite{zvr}    & DE\cite{de}     &EFG\cite{efg}    &LNR\cite{Lahde:1999ih}    \\\hline
  $B(1^1S_0)$      & 5280       & 5280          & 5279            & 5280            & 5277                     \\
  $B(1^3S_1)$      & 5329       & 5330          & 5324            & 5326            & 5325                     \\
  $B(2^1S_0)$      & 5910       & 5830          & 5886            & 5890            & 5822                     \\
  $B(2^3S_1)$      & 5939       & 5870          & 5920            & 5906            & 5848                     \\
  $B(3^1S_0)$      & 6369       & 6210          & 6320            & 6379            & 6117                     \\
  $B(3^3S_1)$      & 6391       & 6240          & 6347            & 6387            & 6136                     \\
  $B(1^3P_0)$      & 5683       & 5650          & 5706            & 5749            & 5678                     \\
  $B_1(1P)$        & 5729       & 5690          & 5700            & 5723            & 5686                     \\
  $B^\prime_1(1P)$ & 5754       & 5690          & 5742            & 5774            & 5699                     \\
  $B(1^3P_2)$      & 5768       & 5710          & 5714            & 5741            & 5704                     \\
  $B(2^3P_0)$      & 6145       & 6060          & 6163            & 6221            & 6010                     \\
  $B_1(2P)$        & 6185       & 6100          & 6175            & 6209            & 6022                     \\
  $B^\prime_1(2P)$ & 6241       & 6100          & 6194            & 6281            & 6028                     \\
  $B(2^3P_2)$      & 6253       & 6120          & 6188            & 6260            & 6040                     \\
  $B(1^3D_1)$      & 6095       & 5970          & 6025            & 6119            & 6005                     \\
  $B_2(1D)$        & 6004       & 5960          & 5985            & 6103            & 5920                     \\
  $B^\prime_2(1D)$ & 6113       & 5980          & 6037            & 6121            & 5955                     \\
  $B(1^3D_3)$      & 6014       & 5970          & 5993            & 6091            & 5871                     \\
  $B(2^3D_1)$      & 6497       & $-$           & $-$             & 6534            & 6248                     \\
  $B_2(2D)$        & 6435       & 6310          & $-$             & 6528            & 6179                     \\
  $B^\prime_2(2D)$ & 6513       & 6320          & $-$             & 6554            & 6207                     \\
  $B(2^3D_3)$      & 6444       & 6320          & $-$             & 6542            & 6140                     \\
  $B(1^3F_2)$      & 6383       & 6190          & 6264            & 6412            & $-$                      \\
  $B_3(1F)$        & 6236       & 6180          & 6220            & 6391            & $-$                      \\
  $B^\prime_3(1F)$ & 6393       & 6200          & 6271            & 6420            & $-$                      \\
  $B(1^3F_4)$      & 6243       & 6180          & 6226            & 6380            & $-$                      \\
  \hline\hline

\end{tabular}
\end{center}
\end{table}

\begin{table}[htbp]
\begin{center}
\caption{ \label{bsmass} The $B_s$ meson masses in MeV from
different quark models. The mixing angles of $B_{sL}-B^\prime_{sL}$ obtained in
this work are $\theta^\prime_{1P}=-34.9^\circ$, $\theta^\prime_{2P}=-36.1^\circ$, $\theta^\prime_{1D}=-39.8^\circ$,
$\theta^\prime_{2D}=-39.8^\circ$, $\theta^\prime_{1F}=-41.1^\circ$. A dash denotes
that the corresponding mass was not calculated in the corresponding reference.}
\footnotesize
\begin{tabular}{lccccc}
\hline\hline
  State            & This work     & ZVR\cite{zvr}  & DE\cite{de}     &EFG\cite{efg}    &LNR\cite{Lahde:1999ih}    \\\hline
  $B_s(1^1S_0)$        & 5362       & 5370          & 5373            & 5372            & 5366                     \\
  $B_s(1^3S_1)$        & 5413       & 5430          & 5421            & 5414            & 5417                     \\
  $B_s(2^1S_0)$        & 5977       & 5930          & 5985            & 5976            & 5939                     \\
  $B_s(2^3S_1)$        & 6003       & 5970          & 6019            & 5992            & 5966                     \\
  $B_s(3^1S_0)$        & 6415       & 6310          & 6421            & 6467            & 6254                     \\
  $B_s(3^3S_1)$        & 6435       & 6340          & 6449            & 6475            & 6274                     \\
  $B_s(1^3P_0)$        & 5756       & 5750          & 5804            & 5833            & 5781                     \\
  $B_{s1}(1P)$         & 5801       & 5790          & 5805            & 5831            & 5795                     \\
  $B^\prime_{s1}(1P)$  & 5836       & 5800          & 5842            & 5865            & 5805                     \\
  $B_s(1^3P_2)$        & 5851       & 5820          & 5820            & 5842            & 5815                     \\
  $B_s(2^3P_0)$        & 6203       & 6170          & 6264            & 6318            & 6143                     \\
  $B_{s1}(2P)$         & 6241       & 6200          & 6278            & 6321            & 6153                     \\
  $B^\prime_{s1}(2P)$  & 6297       & 6210          & 6296            & 6345            & 6160                     \\
  $B_s(2^3P_2)$        & 6309       & 6220          & 6292            & 6359            & 6170                     \\
  $B_s(1^3D_1)$        & 6142       & 6070          & 6127            & 6209            & 6094                     \\
  $B_{s2}(1D)$         & 6087       & 6070          & 6095            & 6189            & 6043                     \\
  $B^\prime_{s2}(1D)$  & 6159       & 6080          & 6140            & 6218            & 6067                     \\
  $B_s(1^3D_3)$        & 6096       & 6080          & 6103            & 6191            & 6016                     \\
  $B_s(2^3D_1)$        & 6527       & $-$           & $-$             & 6629            & 6362                     \\
  $B_{s2}(2D)$         & 6492       & 6410          & $-$             & 6625            & 6320                     \\
  $B^\prime_{s2}(2D)$  & 6542       & 6420          & $-$             & 6651            & 6339                     \\
  $B_s(2^3D_3)$        & 6500       & 6420          & $-$             & 6637            & 6298                     \\
  $B_s(1^3F_2)$        & 6412       & 6300          & 6369            & 6501            & $-$                      \\
  $B_{s3}(1F)$         & 6313       & 6280          & 6332            & 6468            & $-$                      \\
  $B^\prime_{s3}(1F)$  & 6422       & 6310          & 6376            & 6515            & $-$                      \\
  $B_s(1^3F_4)$        & 6319       & 6290          & 6337            & 6475            & $-$                      \\
  \hline\hline

\end{tabular}
\end{center}
\end{table}

\begin{figure*}[htbp]
\includegraphics[scale=0.55]{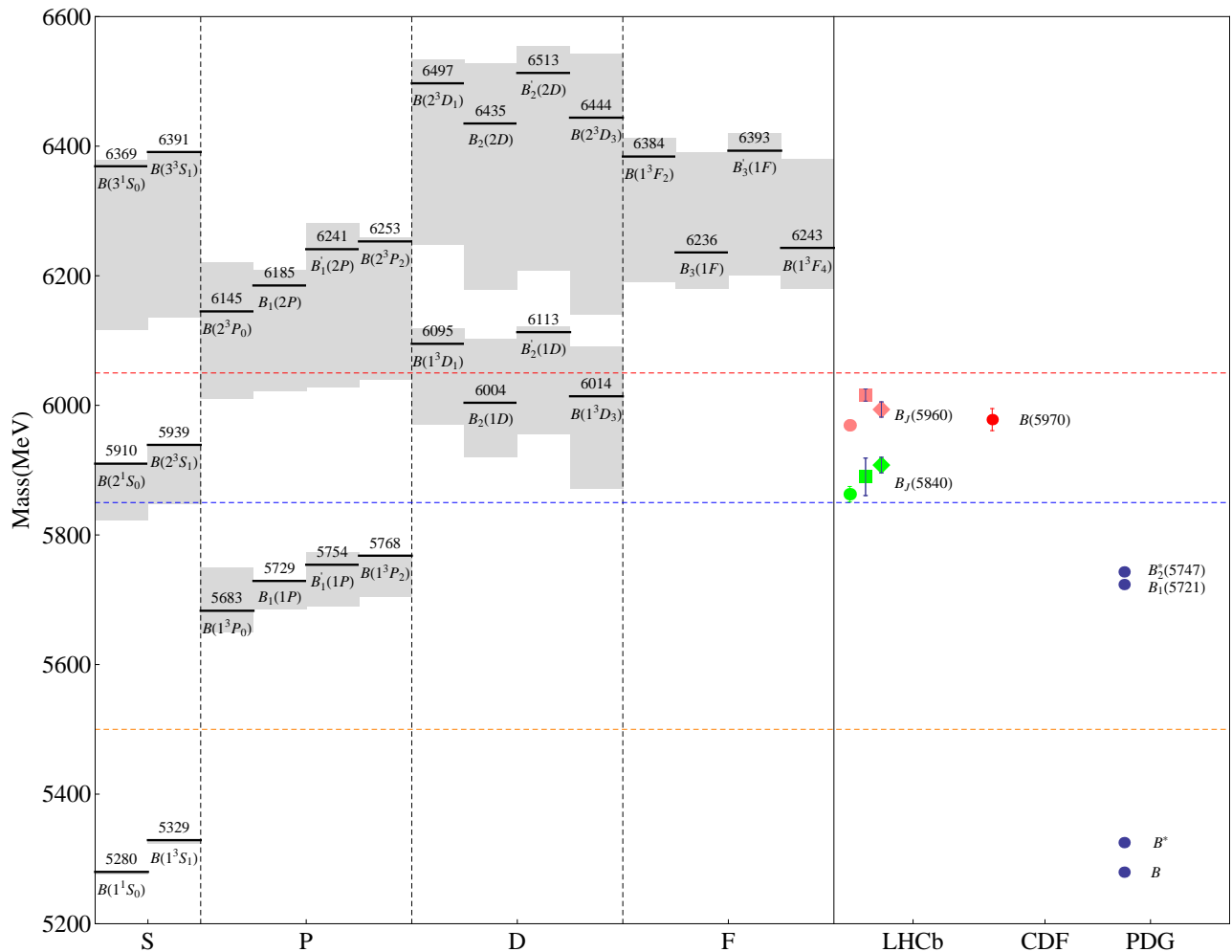}
\vspace{0.0cm} \caption{The bottom meson spectrum. The solid lines stand for our predictions and the
shaded regions are the expected mass ranges from some other quark models\cite{zvr,de,efg,Lahde:1999ih}.
The observed bottom states are also depicted. The N and UN denote natural parity and unnatural parity, respectively.}
\label{fbmass}%
\end{figure*}

\begin{table}
\begin{center}
\caption{\label{assignment} Possible assignments for the observed bottom and bottom-strange states based on masses and spin-parity.} \footnotesize
\begin{tabular}{ll}
\hline\hline
 State                       &  Possible assignments  \\\hline
$B_1(5721)$                  & $B_1(1P)$, $B^\prime_1(1P)$          \\
$B_2^*(5747)$                & $B(1^3P_2)$   \\
$B(5970)$                    & $B(2^3S_1)$, $B(1^3D_3)$ \\
\hline
$B_J(5840)$UN                & $B(2^1S_0)$   \\
$B_J(5960)$UN                & $B_2(1D)$ \\
\hline
$B_J(5840)$N                 & $B(2^3S_1)$   \\
$B_J(5960)$UN                & $B_2(1D)$ \\
\hline
$B_J(5840)$UN                & $B(2^1S_0)$   \\
$B_J(5960)$N                 & $B(2^3S_1)$, $B(1^3D_3)$ \\
\hline
$B_{s1}(5830)$             & $B_{s1}(1P)$, $B^\prime_{s1}(1P)$  \\
$B_{s2}^*(5840)$            & $B_s(1^3P_2)$ \\
\hline
\hline\hline
\end{tabular}
\end{center}
\end{table}

It is believed that the $q\bar{q}$ system can be described by the relativistic Hamiltonian
\begin{eqnarray}
H=\sqrt{p^2+m^2_q}+\sqrt{p^2+m^2_{\bar{q}}}+V(r).
\label{relh}
\end{eqnarray}
For the heavy quarkonium, because of $p$ being smaller than the quark mass, the kinetic energy terms in Eq. (\ref{relh}) can be expanded in inverse powers of the quark mass to obtain the nonrelativistic Hamiltonian. The heavy-light meson state is in principle a relativistic system since
 $p$ is not smaller than the light constituent quark mass $m$. Therefore, it is necessary to discuss the applicability of the nonrelativistic approximation for the relativistic heavy-light system. It is suggested that the following Martin's operator bound is valid for arbitrary mass $M$\cite{Martin:1988zm}
\begin{eqnarray}
\sqrt{p^2+m^2}\leq \frac{M}{2}+\frac{p^2}{2M}+\frac{m^2}{2M}.
\label{mbound}
\end{eqnarray}
This equation gives an upper limit of the relativistic kinetic energy term. The equality in Eq. (\ref{mbound}) holds if the extremum is taken in M. Based on this effective mass expansion of the relativistic kinetic energy term, Jaczko and Durand explain the success of Martin's nonrelativistic descriptions for the spectra of the relativistic light-light and heavy-light mesons\cite{Jaczko:1998uj}. In fact, when an extremum is taken already in the spectrum, the resulting procedure is just a variational method, which establishes a connection between
the nonrelativistic potential model and the relativistic potential model. This method, based on the original work\cite{Brink:1976uf} and also known as the auxiliary or einbein field method, has proven to be rather accurate in various calculations for the relativistic systems\cite{Kalashnikova:2001ig} and has been applied to
the light-light and heavy-light mesons, glueballs, and hybrids\cite{Kalashnikova:2001ig,Simonov:1989fd,Kalashnikova:1995zw,Kaidalov:1999de,
Kalashnikova:2001px,Simonov:2004rh,Buisseret:2006wc}.
This suggests that one can describe the relativistic heavy-light mesons with formally nonrelativistic formulae.

When we discuss the possible assignments of the observed bottom states based on the mass information, we use the mass ranges from different quark models including the nonrelativistic and relativistic models rather than only from the nonrelativistic model (\ref{Ham}).

The predicted mass ranges from different quark models and the
observed bottom states are shown in Fig.~\ref{fbmass}. It is shown
that the ground states $B$ and $B^*$ can be well described. For the
$P$ wave bottom mesons, $B_1(5721)$ and $B_2^*(5747)$ lie within
the $1P$ mass range. Hence, The $B_1(5721)$ can be regarded as
$B_1(1P)$ or $B^\prime_1(1P)$, and $B_2^*(5747)$ is identified as
$B(1^3P_2)$. The observed $B_J(5840)$, $B_J(5960)$, and $B(5970)$
lie close to
 the mass ranges of the $B(2^1S_0)$, $B(2^3S_1)$, $B_2(1D)$, and $B(1^3D_3)$ states.  Considering the spin-parity and masses, we tentatively identify $B(5970)$ as the $B(2^3S_1)$ or $B(1^3D_3)$ state. For the $B_J(5840)$ and $B_J(5960)$, all spin-parity hypotheses should be considered. The bottom-strange mesons $B_s$ and $B_s^*$ are well established. $B_{s1}(5830)$ and $B_{s2}^*(5840)$ can be clarified into $P$ wave bottom-strange mesons. The assignments for these observed bottom and bottom-strange states are listed in Table \ref{assignment}. Below, we shall focus on these possible assignments.
Since the mass information alone is insufficient to identify these states, hence the strong decay behaviors also need to be investigated in the $^3P_0$ model.

\section{Strong decays}

\subsection{$^3P_0$ model}

In this work, we adopt the $^3P_0$ model to evaluate the
Okubo-Zweig-Iizuka-allowed two-body strong decays of the bottom and bottom-strange mesons. The $^3P_0$ model, also called as quark pair creation model, has been wildly applied to study hadron strong decays with considerable success\cite{3p0model1,3p0model2,3p0model3,3p0model4}. In this model, the meson decay occurs through a quark-antiquark pair with the vacuum quantum number\cite{micu}. Here we give a brief review of the $^3P_0$ model. The transition operator $T$ of the decay $A\rightarrow BC$ in the
$^3P_0$ model can be written as\cite{3p0decay}
\begin{eqnarray}
T=-3\gamma\sum_m\langle 1m1-m|00\rangle\int
d^3\boldsymbol{p}_3d^3\boldsymbol{p}_4\delta^3(\boldsymbol{p}_3+\boldsymbol{p}_4)\nonumber\\
{\cal{Y}}^m_1\left(\frac{\boldsymbol{p}_3-\boldsymbol{p}_4}{2}\right
)\chi^{34}_{1-m}\phi^{34}_0\omega^{34}_0b^\dagger_3(\boldsymbol{p}_3)d^\dagger_4(\boldsymbol{p}_4),
\end{eqnarray}
where $\gamma$ is a dimensionless $q_3\bar{q}_4$ pair-production strength, and $\boldsymbol{p}_3$ and
$\boldsymbol{p}_4$ are the momenta of the created quark $q_3$ and
antiquark  $\bar{q}_4$, respectively. $\phi^{34}_{0}$,
$\omega^{34}_0$, and $\chi_{{1,-m}}^{34}$ are the flavor, color,
and spin wave functions of the  $q_3\bar{q}_4$, respectively. The
solid harmonic polynomial
${\cal{Y}}^m_1(\boldsymbol{p})\equiv|p|^1Y^m_1(\theta_p, \phi_p)$ reflects
the momentum-space distribution of the $q_3\bar{q}_4$ .

The partial wave amplitude ${\cal{M}}^{LS}(\boldsymbol{P})$ can be expressed as
\begin{eqnarray}
{\cal{M}}^{LS}(\boldsymbol{P})&=&
\sum_{\renewcommand{\arraystretch}{.5}\begin{array}[t]{l}
\scriptstyle{M_{J_B},M_{J_C},M_S,M_L}
\end{array}}\renewcommand{\arraystretch}{1}\!\!
\langle LM_LSM_S|J_AM_{J_A}\rangle \nonumber\\
&&\times\langle
J_BM_{J_B}J_CM_{J_C}|SM_S\rangle\nonumber\\
&&\times\int
d\Omega\,\mbox{}Y^\ast_{LM_L}{\cal{M}}^{M_{J_A}M_{J_B}M_{J_C}}
(\boldsymbol{P}), \label{pwave}
\end{eqnarray}
where ${\cal{M}}^{M_{J_A}M_{J_B}M_{J_C}}
(\boldsymbol{P})$ is the helicity amplitude and defined as
\begin{eqnarray}
\langle
BC|T|A\rangle=\delta^3(\boldsymbol{P}_A-\boldsymbol{P}_B-\boldsymbol{P}_C){\cal{M}}^{M_{J_A}M_{J_B}M_{J_C}}(\boldsymbol{P}).
\end{eqnarray}
The $|A\rangle$, $|B\rangle$, and $|C\rangle$ denote the mock meson states and the mock meson $|A\rangle$
is defined by\cite{mockmeson}
\begin{eqnarray*}
&&|A(n^{2S_A+1}_AL_{A}\,\mbox{}_{J_A M_{J_A}})(\boldsymbol{P}_A)\rangle
\equiv \nonumber\\
&& \sqrt{2E_A}\sum_{M_{L_A},M_{S_A}}\langle L_A M_{L_A} S_A
M_{S_A}|J_A
M_{J_A}\rangle\nonumber\\
&&\times  \int d^3\boldsymbol{p}_A\psi_{n_AL_AM_{L_A}}(\boldsymbol{p}_A)\chi^{12}_{S_AM_{S_A}}
\phi^{12}_A\omega^{12}_A\nonumber\\
&&\times  \left|q_1\left({\scriptstyle
\frac{m_1}{m_1+m_2}}\boldsymbol{P}_A+\boldsymbol{p}_A\right)\bar{q}_2
\left({\scriptstyle\frac{m_2}{m_1+m_2}}\boldsymbol{P}_A-\boldsymbol{p}_A\right)\right\rangle,
\end{eqnarray*}
where $m_1$ and $m_2$ ($\boldsymbol{p}_1$ and $\boldsymbol{p}_2$) are the masses (momenta) of the
quark $q_1$ and the antiquark $\bar{q}_2$, respectively; $\boldsymbol{P}_A=\boldsymbol{p}_1+\boldsymbol{p}_2$,
$\boldsymbol{p}_A=\frac{m_2\boldsymbol{p}_1-m_1\boldsymbol{p}_2}{m_1+m_2}$;
$\chi^{12}_{S_AM_{S_A}}$, $\phi^{12}_A$, $\omega^{12}_A$,
$\psi_{n_AL_AM_{L_A}}(\boldsymbol{p}_A)$ are the spin, flavor, color, and
space wave functions of the meson $A$ composed of $q_1\bar{q}_2$ with total energy $E_A$, respectively.

Because of different choices of the pair-production vertex, phase space conventions, employed meson wave functions, various $^3P_0$ models exist in literatures. In this work, we restrict to the simplest vertex as introduced originally by\cite{micu}, which assumes a
spatially constant pair-production strength $\gamma$, adopt the relativistic phase space as Ref.\cite{3p0decay}, and the realistic meson wave functions from the quark model(\ref{Ham}).
With the relativistic phase space, the decay width
$\Gamma(A\rightarrow BC)$ can be expressed in terms of the partial wave
amplitude Eq.~(\ref{pwave})
\begin{eqnarray}
\Gamma(A\rightarrow BC)= \frac{\pi
P}{4M^2_A}\sum_{LS}|{\cal{M}}^{LS}(\boldsymbol{P})|^2, \label{width1}
\end{eqnarray}
where $P=|\boldsymbol{P}|=\frac{\sqrt{[M^2_A-(M_B+M_C)^2][M^2_A-(M_B-M_C)^2]}}{2M_A}$,
and $M_A$, $M_B$, and $M_C$ are the masses of the mesons $A$, $B$,
and $C$, respectively.

We take the light nonstrange quark pair creation strength $\gamma =7.6$ by fitting to the total width of the $B_2^*(5747)$ as the $B(1^3P_2)$ state. The $\gamma$ and strange quark pair creation
  strength $\gamma_{s\bar{s}}$ can be related by $\gamma_{s\bar{s}}=\gamma\frac{m_u}{m_s}$\cite{rss}, where the constituent quark masses $m_u$ and $m_s$
  the same as those used in the mass esitmateds in the quark model (\ref{Ham}). Our value of $\gamma$ is higher than that used by other groups such as\cite{Close:2005se, 3p0model4} by a factor of $\sqrt{96\pi}$ due to different field conventions. The mixing angles $\theta_{nL}$ are taken from Table \ref{bmass} and \ref{bsmass}.

\subsection{$B(1P)$ states}

For the $B(1^3P_0)$ state, the predicted mass is above the $B\pi$ threshold. The decay widths of the $B(1^3P_0)$ are shown in Table~\ref{B13P0}. No experimental data of the $B(1^3P_0)$ exist, but some theoretical estimations also give a broad width\cite{Zhong:2010vq,Sun:2014wea}, which is consistent with our result.

\begin{table}
\begin{center}
\caption{ \label{B13P0} Decay widths of the $B(1^3P_0)$ in MeV.}
\footnotesize
\begin{tabular}{lc}
\hline\hline
  $B^+\pi^-$                                 & 153.80\\
  $B^0\pi^0$                                 & 76.62\\
  Total width                                & 230.43 \\
\hline\hline
\end{tabular}
\end{center}
\end{table}

In Table \ref{B13P2}, we give the decay widths of the the $B_2^*(5747)$.  The $\gamma$-independent ratio is predicted to be
\begin{eqnarray}
\frac{\Gamma(B_2^*(5747)^0\rightarrow B^*+\pi^-)}{\Gamma(B_2^*(5747)^0\rightarrow B^{+}\pi^-)}=0.95,
\end{eqnarray}
which is consistent with experimental data of $1.10\pm 0.42\pm
0.31$\cite{Abazov:2007vq} and $0.71\pm 0.14\pm
0.30$\cite{Aaij:2015qla}.

\begin{table}
\begin{center}
\caption{ \label{B13P2} Decay widths of the $B_2^*(5747)$ in MeV.}
\footnotesize
\begin{tabular}{lc}
\hline\hline
  $B^+\pi^-$                                 & 8.46\\
  $B^0\pi^0$                                 & 4.16\\
  $B^{*+}\pi^-$                              & 7.97\\
  $B^{*0}\pi^0$                              & 3.92\\
  Total width                                & 24.51 \\
  Experiment                                 & $24.5\pm1.0\pm11.5$\\
\hline\hline
\end{tabular}
\end{center}
\end{table}

The decay widths of the $B_1(5721)$ as the $B_1(1P)$ and
$B^\prime_1(1P)$ are listed in Table \ref{B1P}. With the $B_1(1P)$ assignment to $B_1(5721)$, the total decay width
is expected to be about 200 MeV, much larger than the experiment, hence this assignment can be totally excluded. With the $B^\prime_1(1P)$ assignment, the total width of the $B_1(5721)$ is 40.63 MeV, consistent with $30.1\pm1.5\pm3.5$ given by the LHCb Collaboration\cite{Aaij:2015qla}.

\begin{table}
\begin{center}
\caption{ \label{B1P} Decay widths of the $B_1(5721)$ in MeV.}
\footnotesize
\begin{tabular}{lcc}
\hline\hline
 Channel                        & $B_1(1P)$     & $B^\prime_1(1P)$ \\
\hline
  $B^{*+}\pi^-$                 & 132.77        &27.12  \\
  $B^{*0}\pi^0$                 & 66.63         &13.51   \\
  Total width                   & 199.40        &40.63   \\
  Experiment                    &\multicolumn{2}{c}{$30.1\pm1.5\pm3.5$}  \\
\hline\hline
\end{tabular}
\end{center}
\end{table}

\begin{figure}[htbp]
\includegraphics[scale=0.7]{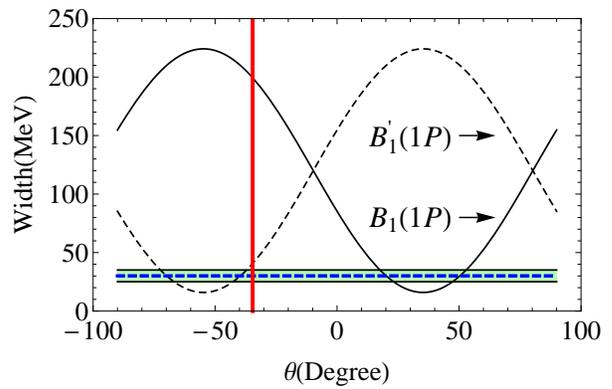}
\vspace{0.0cm} \caption{Total decay width of the $B_1(5721)$ as
the $B_1(1P)$ and $B^\prime_1(1P)$ versus the mixing angle. The blue dashed line with a
green band denotes the LHCb experimental data. The vertical red solid line corresponds to
the mixing angle $\theta_{1P}=-34.6^\circ$ obtained in Sec.~\ref{masses}.}
\label{fB1P}
\end{figure}

The dependence of the $B_1(5721)$ total width on the mixing angle $\theta_{1P}$ is depicted in Fig \ref{fB1P}. The predicted mixing angle from the quark model (\ref{Ham}) is $\theta_{1P} = -34.6^\circ$. With this angle, the $B_1(1P)$ decay width is much broader than that of the $B^\prime_1(1P)$, which is consistent with other theoretical predictions\cite{Zhong:2010vq,Sun:2014wea}. In the heavy quark effective theory, the $P$ wave heavy-light mesons can be divided into the $(0^+,1^+)_{j=\frac{1}{2}}$ and $(1^+,2^+)_{j=\frac{3}{2}}$ doublets, where $j$ is the total angular momentum of the light quark. In the heavy quark limit, the $\theta_{1P} = -54.7^\circ$, which is close to quark model prediction of $-34.6^\circ$. For the two $1^+$ states, the decay width is broader for the $j=\frac{1}{2}$ state than that for the $j=\frac{3}{2}$ state. Hence, with the $B^\prime_1(1P)$ assignment, the $B_1(5721)$ corresponds to the $1^+$ bottom meson belonging to the $(1^+,2^+)_{j=\frac{3}{2}}$ doublets.

\subsection{$B_J(5840)$ and $B_J(5960)$}

For the $B_J(5840)$ and $B_J(5960)$, three spin-parity hypothesis exist, which classifies these states into different possible assignments. In the following, we will consider these assignments one by one.

In Table \ref{B5840a}, we list the decay widths of the $B_J(5840)$ and $B_J(5960)$ under the hypothesis I.
The predicted total width of the $B_J(5840)$ as the $B(2^1S_0)$
is 126.22 MeV, in good agreement with the experimental data of $127.4\pm16.7\pm34.2$ MeV. However, the predicted total width of the $B_J(5960)$ as the $B_2(1D)$ is much larger than the experimental data of $82.3\pm7.7\pm9.4$ MeV.

\begin{table}
\begin{center}
\caption{ \label{B5840a} Decay widths of the $B_J(5840)$ and $B_J(5960)$ under the hypothesis I in MeV. A dash indicates that a decay mode is forbidden.}
\scriptsize
\begin{tabular}{lccc}
\hline\hline
                        &$B_J(5840)$         & $B_J(5960)$      \\
                        &$B(2^1S_0)$         & $B_2(1D)$      \\\hline
  $B^{*+}\pi^-$         & 84.13              & 63.51             \\
  $B^{*0}\pi^0$         & 42.09               & 31.83                 \\
  $B(1^3P_0)^+\pi^-$    & $-$               & 69.62                  \\
  $B(1^3P_0)^0\pi^0$    & $-$               & 34.61                     \\
  $B^*_2(5747)^+\pi^-$  & $-$                & 0.005                   \\
  $B^*_2(5747)^0\pi^0$  & $-$                & 0.002                   \\
  $B_1(1P)^+\pi^-$      & $-$               & 0.01                     \\
  $B_1(1P)^0\pi^0$      & $-$               & 0.005                        \\
  $B^\prime_1(1P)^+\pi^-$    & $-$         & 0.17                     \\
  $B^\prime_1(1P)^+\pi^-$    & $-$         & 0.08                          \\
  $B^*\eta$             & $-$               & 11.41                        \\
  $B_s^*K$              & $-$              & 10.98                         \\
  Total width           & 126.22             & 222.24               \\
  Experiment            & $127.4\pm16.7\pm34.2$   & $82.3\pm7.7\pm9.4$
 \\
\hline\hline
\end{tabular}
\end{center}
\end{table}

The predicted decay widths of the $B_J(5840)$ and $B_J(5960)$ under the hypothesis II are presented in Table \ref{B5840b}. The predicted total width of the $B_J(5840)$ as the $B(2^3S_1)$ state
is 106.13 MeV, consistent with the experimental data of $107.0\pm19.7\pm34.2$ MeV. However, the predicted total width of the $B_J(5960)$ as the $B_2(1D)$ state is much larger than the experimental data of $81.6\pm9.9\pm9.4$ MeV.

\begin{table}
\begin{center}
\caption{ \label{B5840b} Decay widths of the $B_J(5840)$ and $B_J(5960)$ under the hypothesis II in MeV. A dash indicates that a decay mode is forbidden.}
\scriptsize
\begin{tabular}{lccc}
\hline\hline
                        &$B_J(5840)$         & $B_J(5960)$      \\
                        &$B(2^3S_1)$         & $B_2(1D)$    \\\hline
  $B^+\pi^-$            & 20.43             & $-$            \\
  $B^0\pi^0$            & 10.26             & $-$                     \\
  $B^{*+}\pi^-$         & 46.68              & 59.21                     \\
  $B^{*0}\pi^0$         & 23.39               & 29.70                       \\
  $B(1^3P_0)^+\pi^-$    & 0.002               & 84.63                     \\
  $B(1^3P_0)^0\pi^0$    & $2.9\times10^{-4}$  & 42.42                          \\
  $B^*_2(5747)^+\pi^-$  & $-$                & 0.01                 \\
  $B^*_2(5747)^0\pi^0$  & $-$                & 0.006                 \\
  $B_1(1P)^+\pi^-$      & 0.21               & 0.05                      \\
  $B_1(1P)^0\pi^0$      & 0.10              & 0.02                          \\
  $B^\prime_1(1P)^+\pi^-$    & 0.05         & 0.54                    \\
  $B^\prime_1(1P)^+\pi^-$    & 0.02         & 0.26                         \\
  $B\eta$               & 2.61              & $-$                     \\
  $B^*\eta$             & 0.98              & 16.79                      \\
  $B_sK$                & 1.40              & $-$                      \\
  $B_s^*K$              & $-$               & 24.14                        \\
  Total width           & 106.13             & 257.79                \\
  Experiment            & $107.0\pm19.6\pm34.2$   & $81.6\pm9.9\pm9.4$
 \\
\hline\hline
\end{tabular}
\end{center}
\end{table}

The decay widths of the $B_J(5840)$ and $B_J(5960)$ under the hypothesis III are shown in Table \ref{B5840c}.
The predicted width of the $B_J(5840)$ as the $B(2^1S_0)$ is about 134 MeV, consistent with
 the measured result of $119.4\pm17.2\pm34.2$ MeV. If the $B_J(5960)$ is the $B(2^3S_1)$, the $B_J(5960)$ total width is expected to be $131.97$ MeV, far away from the measured width of $55.9\pm6.6\pm9.4$ MeV. If the $B_J(5960)$ is the $B(1^3D_3)$, the $B_J(5960)$ total width is expected to be 48.55 MeV, in good agreement with the data of $55.9\pm 6.6\pm 9.4$ MeV.

\begin{table}
\begin{center}
\caption{ \label{B5840c} Decay widths of the $B_J(5840)$ and $B_J(5960)$ under the hypothesis III in MeV. A dash indicates that a decay mode is forbidden.}
\scriptsize
\begin{tabular}{lcccc}
\hline\hline
                         &$B_J(5840)$           &\multicolumn{2}{c} {$B_J(5960)$}    \\
                         &$B(2^1S_0)$             &$B(2^3S_1)$         & $B(1^3D_3)$     \\\hline
  $B^+\pi^-$             & $-$                   & 14.39              & 15.08            \\
  $B^0\pi^0$             & $-$                   & 7.25               & 7.48              \\
  $B^{*+}\pi^-$          & 86.24                 & 38.20              & 16.17             \\
  $B^{*0}\pi^0$          & 43.21                 & 19.20              & 8.03               \\
  $B^{*+}(1^3P_0)\pi^-$  & 4.1$\times10^{-4}$    & $-$                & $-$                \\
  $B^{*0}(1^3P_0)\pi^0$  & 4.0$\times10^{-5}$    & $-$                & $-$                \\
  $B^*_2(5747)^+\pi^-$   & 0.05                  & 1.62               & 0.35               \\
  $B^*_2(5747)^0\pi^0$   & 0.02                  & 0.76               & 0.16                \\
  $B_1(1P)^+\pi^-$       & $-$                   & 0.37               & 0.15                \\
  $B_1(1P)^0\pi^0$       & $-$                   & 0.18               & 0.07                \\
  $B^\prime_1(1P)^+\pi^-$ & $-$                  & 2.41               & 0.04                \\
  $B^\prime_1(1P)^+\pi^-$ & $-$                  & 1.14               & 0.02                \\
  $B\eta$                & $-$                   & 6.74               & 0.44                \\
  $B^*\eta$              & 4.57                  & 12.06              & 0.23                \\
  $B_sK$                 & $-$                   & 11.81               & 0.26                \\
  $B_s^*K$               & $-$                   & 15.83              & 0.08                \\
  Total width            & 134.09                & 131.97             & 48.55               \\
  Experiment             & $119.4\pm17.2\pm34.2$ &\multicolumn{2}{c} {$55.9\pm6.6\pm9.4$}    \\
\hline\hline
\end{tabular}
\end{center}
\end{table}

To sum up, with the hypothesis III, the total widths of the $B_J(5840)$ and $B_J(5960)$ can be reproduced simultaneously. The strong decay behaviors combined with masses indicate that the $B_J(5840)$ and $B_J(5960)$ can be identified as the $B(2^1S_0)$ and $B(1^3D_3)$, respectively. The assignment of the $B_J(5840)$ as the $B(2^1S_0)$ state, is also suggested by the LHCb Collaboration\cite{Aaij:2015qla}. The main decay modes of the $B(2^1S_0)$ are expected to $B^*\pi$ and $B^*\eta$. The main decay modes of the $B(1^3D_3)$ are expected to be $B\pi$ and $B^*\pi$.

\subsection{$B(5970)$}

The decay widths of the $B(5970)$ as the $B(2^3S_1)$ and
$B(1^3D_3)$ are listed in Table \ref{B5970}. Since the $B(5970)$ mass is close to the $B_J(5960)$ mass,
the results for the $B(2^3S_1)$ and $B(1^3D_3)$ are similar with those in Table \ref{B5840c}. However,
  because of the large uncertainty of the $B(5970)$ total width, both the $B(2^3S_1)$ and $B(1^3D_3)$ assignments
  are favored by the experimental data\cite{Aaltonen:2013atp}. Ref.\cite{Sun:2014wea} interprets the $B(5870)$ as the $B(2^3S_1)$ state while Ref.\cite{Zhong:2010vq} assigns  the $B(5970)$ as the $B(1^3D_3)$ state\cite{Zhong:2010vq}. The main decay modes of the $B(2^3S_1)$ are expected to be $B\pi$, $B^*\pi$, $B\eta$, $B^*\eta$, $B_sK$, and $B^*_sK$, while the $B(1^3D_3)$ is expected to mainly decay to $B\pi$, $B^*\pi$. Further precise measurements of the width, spin and decay modes are needed to distinguish these two assignments.

\begin{table}
\begin{center}
\caption{ \label{B5970} Decay widths of the $B(5970)$ as the $B(2^3S_1)$ and
$B(1^3D_3)$ in MeV.}
\scriptsize
\begin{tabular}{lcccc}
\hline\hline
                               &\multicolumn{2}{c} {$B(5970)$}\\
                               & $B(2^3S_1)$      & $B(1^3D_3)$     \\\hline
  $B^+\pi^-$                   & 15.54            & 13.496             \\
  $B^0\pi^0$                   & 7.83             & 6.69          \\
  $B^{*+}\pi^-$                & 40.26            & 14.26              \\
  $B^{*0}\pi^0$                & 20.23            & 7.08                \\
  $B^*_2(5747)^+\pi^-$         & 1.02             & 0.21                \\
  $B^*_2(5747)^0\pi^0$         & 0.47             & 0.10                \\
  $B_1(1P)^+\pi^-$             & 0.31             & 0.10                \\
  $B_1(1P)^0\pi^0$             & 0.15             & 0.05                \\
  $B^\prime_1(1P)^+\pi^-$      & 1.63             & 0.02             \\
  $B^\prime_1(1P)^+\pi^-$      & 0.77             & 0.12                \\
  $B\eta$                      & 6.38             & 0.31                \\
  $B^*\eta$                    & 10.66             & 0.14                \\
  $B_sK$                       & 10.35             & 0.16             \\
  $B_s^*K$                     & 12.12            & 0.03              \\
  Total width                  & 127.72             & 42.69       \\
  Experiment                   &\multicolumn{2}{c} {$70^{+30}_{-20}\pm30$}\\
\hline\hline
\end{tabular}
\end{center}
\end{table}

\subsection{$B(1^3D_1)$, $B_2(1D)$ and $B^\prime_2(1D)$}

Given the bottom masses and spin-parity, no experimental candidates exist for the $B(1^3D_1)$, $B_2(1D)$ and $B^\prime_2(1D)$ states. Our predicted masses for these three states are 6095 MeV, 6004 MeV, and 6113 MeV, respectively. With these masses as inputs, their total decay widths are listed in Table \ref{B1D}.

\begin{table}
\begin{center}
\caption{ \label{B1D} Decay widths of the $B(1^3D_1)$, $B_2(1D)$ and $B^\prime_2(1D)$ states in MeV. A dash indicates that a decay mode is forbidden.}
\scriptsize
\begin{tabular}{lccc}
\hline\hline

                        &$B(1^3D_1)$         & $B_2(1D)$   & $B^\prime_2(1D)$      \\\hline
  $B^+\pi^-$            & 26.44              & $-$         & $-$             \\
  $B^0\pi^0$            & 13.31              & $-$        & $-$                \\
  $B^{*+}\pi^-$         & 15.45              & 60.57           & 61.08                \\
  $B^{*0}\pi^0$         & 7.76               & 30.38           & 30.44                  \\
  $B(1^3P_0)^+\pi^-$    & 4.52               & 81.79          & 3.51                 \\
  $B(1^3P_0)^0\pi^0$    & 2.21               & 40.93          & 1.76                     \\
  $B^*_2(5747)^+\pi^-$  & $-$                & 0.01         & 0.20                \\
  $B^*_2(5747)^0\pi^0$  & $-$                & 0.005          & 0.10              \\
  $B_1(1P)^+\pi^-$      & 10.30               & 0.03          & 0.18                 \\
  $B_1(1P)^0\pi^0$      & 5.20               & 0.02          & 0.09                     \\
  $B^\prime_1(1P)^+\pi^-$    & 74.77         & 0.42          & 5.57                 \\
  $B^\prime_1(1P)^+\pi^-$    & 37.71         & 0.20          & 2.74                     \\
  $B\eta$               & 14.18              & $-$          & $-$                   \\
  $B^*\eta$             & 7.34               & 15.58         & 4.64                   \\
  $B_sK$                & 32.67              & $-$          & $-$                  \\
  $B_s^*K$              & 15.06              & 20.75           & 5.04                    \\
  $B^+\rho^-$           & 10.64               & $-$         & 48.37              \\
  $B^0\rho^0$           & 5.27               & $-$         & 24.04                \\
  $B^{*+}\rho^-$        & $-$                & $-$           & 4.02                \\
  $B^{*0}\rho^0$        & $-$                & $-$          & 2.01                  \\
  $B\omega$             & 4.20               & $-$         & 21.08              \\
  $B^*\omega$           & $-$                & $-$           & 0.58                  \\
  $B(2^1S_0)^+\pi^-$    & 0.52               & $-$          & $-$                 \\
  $B(2^1S_0)^0\pi^0$    & 0.24               & $-$          & $-$                     \\
  $B(2^3S_1)^+\pi^-$    & 0.05               & $-$          & 0.01                 \\
  $B(2^3S_1)^0\pi^0$    & 0.02               & $-$          & 0.01                     \\
  Total width           & 287.85             & 250.69        &215.47                 \\
\hline\hline
\end{tabular}
\end{center}
\end{table}

It is shown that all these three states have large total widths more than 200 MeV. The decay modes of these states are different, mainly due to the quantum number conservation and the threshold.

In heavy quark limit, the mixing angle is $\theta_{1D} = -50.8^\circ$\cite{jjcoupling}. Our predicted $\theta_{1D}=-39.6^\circ$ is close to $-50.8^\circ$. With this mixing angle, the total decay width of $B_2(1D)$ is broader than $B^\prime_2(1D)$, which indicate that $B_2(1D)$ and $B^\prime_2(1D)$ corresponds to the $(1^-,2^-)_{j=\frac{3}{2}}$ and $(2^-,3^-)_{j=\frac{5}{2}}$ doublets, respectively.

\subsection{$B_s(1P)$ states}

For the $B_s(1^3P_0)$ state, the predicted mass is below the $BK$ threshold, which is consistent with some other studies\cite{zvr,Cheng:2014bca}. Hence, there is no OZI-allowed strong decay pattern and the dominant decay mode may be $B_s\pi$. This situation is analogous to the charmed-strange partner $D^*_{s0}(2317)$, whose decay width is mainly due to OZI-violated $D_s\pi$ channel. Based on higher mass of $B_s(1^3P_0)$ state, some theoretical calculations give a broad decay width\cite{Zhong:2010vq,Sun:2014wea}. Further experimental search for the $B_s(1^3P_0)$ state will distinguish these two predictions.

The decay widths of the $B_{s2}^*(5840)$ are listed in Table \ref{Bs13P2}. The predicted total decay width is 1.99 MeV, in good agreement with LHCb experimental data of $1.56\pm0.13\pm0.47$ MeV\cite{Agashe:2014kda,Aaij:2012uva} and the CDF result of $1.4\pm0.4\pm0.2$ MeV\cite{Aaltonen:2013atp}. The predicted ratio
\begin{eqnarray}
\frac{\Gamma(B_{s2}^*(5840)\rightarrow B^{*+}K^-)}{\Gamma(B_{s2}^*(5840)\rightarrow B^{+}K^-)}=0.086
\end{eqnarray}
is independent with the $\gamma$ and in agreement with the LHCb experimental data of $0.093\pm 0.013\pm 0.012$\cite{Agashe:2014kda,Aaij:2012uva}.

\begin{table}
\begin{center}
\caption{ \label{Bs13P2} Decay widths of the $B_{s2}^*(5840)$ in MeV.}
\footnotesize
\begin{tabular}{lc}
\hline\hline
  $B^+K^-$                                 & 1.00\\
  $B^0\bar{K}^0$                           & 0.86\\
  $B^{*+}K^-$                              & 0.09\\
  $B^{*0}\bar{K}^0$                        & 0.05\\
  Total width                              & 1.99 \\
  Experiment                               & $1.56\pm0.13\pm0.47$/$1.4\pm0.4\pm0.2$\\
\hline\hline
\end{tabular}
\end{center}
\end{table}

\begin{figure}[htbp]
\includegraphics[scale=0.7]{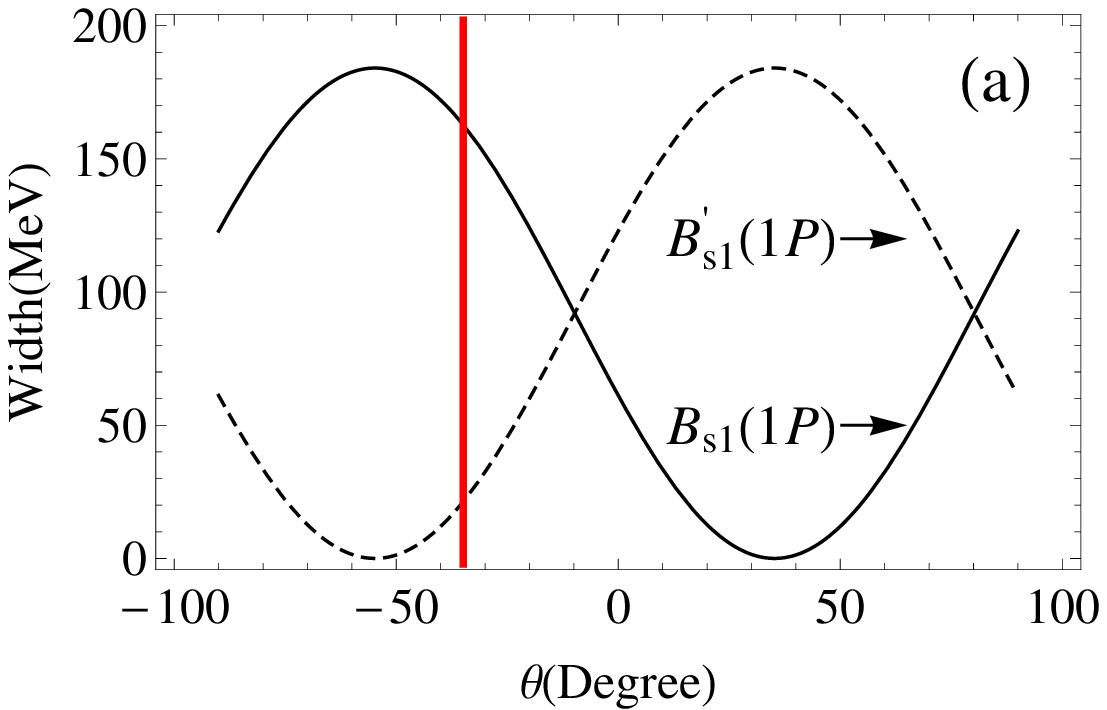}
\includegraphics[scale=0.7]{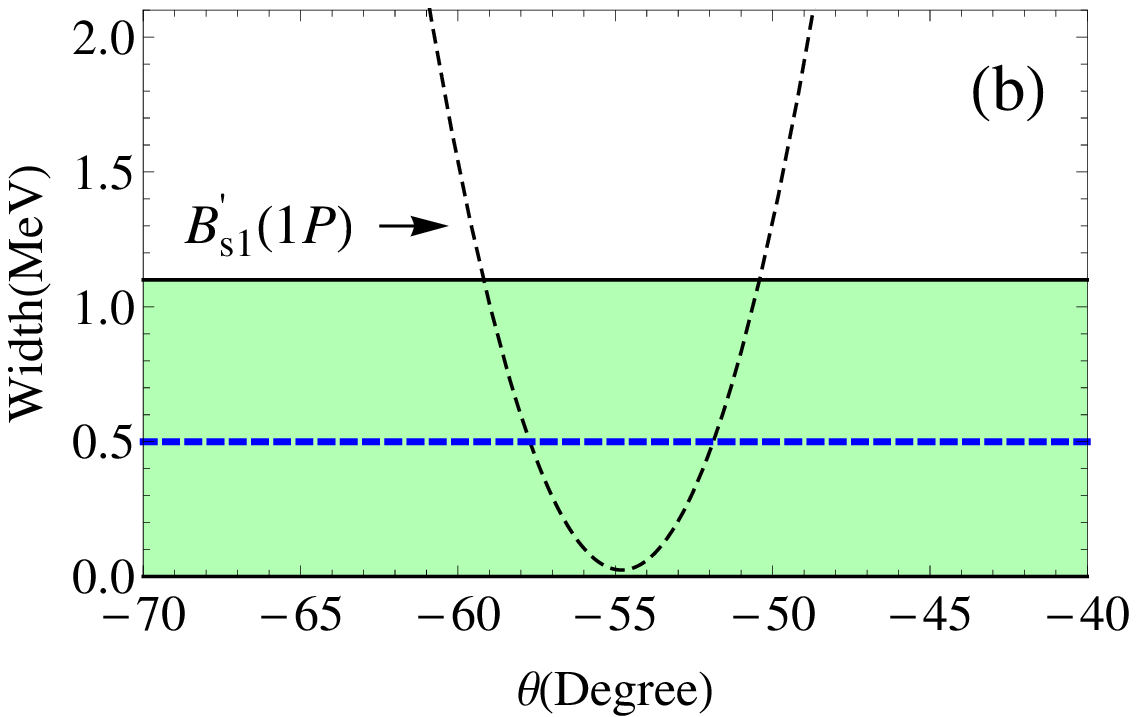}
\vspace{0.0cm} \caption{(a) Total width of the $B_{s1}(5830)$ as
the $B_{s1}(1P)$ and $B^\prime_{s1}(1P)$ versus the mixing angle. The vertical red solid line corresponds to
the mixing angle $\theta_{1P}=-34.9^\circ$ obtained in Sec.~\ref{masses}. (b) The variation of total width of the $B^\prime_{s1}(1P)$ with mixing angle $\theta_{1P}=(-70\sim -40)^\circ$. The blue dashed line with a
green band denotes the CDF experimental data.}
\label{fBs1P}
\end{figure}

In analogous to the $B_1(5721)$, the $B_{s1}(5830)$ can be $B_{s1}(1P)$ or $B^\prime_{s1}(1P)$ . With the predicted mixing angle $\theta_{1P}=-34.9^\circ$, the total widths of $B_{s1}(1P)$ and $B^\prime_{s1}(1P)$ are expected to be  162.76 MeV and 21.35 MeV, respectively, both much lager than the CDF data of $0.5\pm0.3\pm0.3$ MeV\cite{Aaltonen:2013atp}. The dependence of the total widths of $B_{s1}(1P)$ and $B^\prime_{s1}(1P)$ states versus the mixing angle are shown in Fig \ref{fBs1P}. It can be seen that when the mixing angle varies in the range of $(-59.2\sim-50.4)^\circ$, the total width of the $B_{s1}(1P)$ is consistent with the observed width. In the heavy quark effective theory, the ideal value of the mixing angle is $\theta_{1P}=-54.7^\circ$, lying in the range of $(-59.2\sim-50.4)^\circ$. The extremely narrow total width of the $B_{s1}(5830)$ suggests that it can be identified as the $B^\prime_{s1}(1P)$ state belonging to the $(1^+,2^+)_{j=\frac{3}{2}}$ doublet.

\subsection{$B_s(2S)$}

Our predicted masses of the $B_s(2^1S_0)$ and $B_s(2^3S_1)$ are 5977 MeV and 6003 MeV, respectively. The decay widths of the $B_s(2^1S_0)$ and
$B_s(2^3S_1)$ are listed in Table \ref{Bs2S}. It is shown that the $B_s(2^1S_0)$ state mainly decays into $B^*K$, and the main decay modes of the $B_s(2^3S_1)$ state are $BK$ and $B^*K$.

\begin{table}
\begin{center}
\caption{ \label{Bs2S} Decay widths of the $B_s(2^1S_0)$ and
$B_s(2^3S_1)$ in MeV.}
\scriptsize
\begin{tabular}{lcccc}
\hline\hline
                               & $B_s(2^1S_0)$      & $B_s(2^3S_1)$     \\\hline
  $B^+K^-$                     & $-$                & 35.67             \\
  $B^0\bar{K}^0$               & $-$                & 35.67          \\
  $B^{*+}K^-$                  & 106.62             & 70.71              \\
  $B^{*0}\bar{K}^0$            & 105.26             & 70.29                \\
  $B_s\eta$                    & $-$                & 5.32                \\
  $B_s^*\eta$                  & 1.51               & 4.24                \\
  Total width                  & 213.38             & 221.90       \\
\hline\hline
\end{tabular}
\end{center}
\end{table}

\subsection{$B_s(1^3D_1)$, $B_s(1^3D_3)$, $B_{s2}(1D)$, $B^\prime_{s2}(1D)$}

Our predicted masses of the $B_s(1^3D_1)$, $B_s(1^3D_3)$, $B_{s2}(1D)$, $B^\prime_{s2}(1D)$ are
 6142MeV, 6096MeV, 6087 MeV, and 6159MeV, respectively. The decay widths of these states are shown in Table \ref{Bs1D}. The $B_s(1^3D_1)$ state is expected to be broad, while the $B_s(1^3D_3)$ is a narrow state. This behavior is similar with their bottom partners $B(1^3D_1)$ and $B(1^3D_3)$. The dominant decay modes of the $B_s(1^3D_1)$ and $B_s(1^3D_3)$ are $BK$ and $B^*K$.

The mixing angle of the $B_{s2}(1D)$ and $B_{s2}^\prime(1D)$ states is expected to be $-39.8^\circ$, close to the ideal mixing angle $-50.8^\circ$ in the heavy quark limit. For the $B_{s2}(1D)$ and $B_{s2}^\prime(1D)$, the $B^*K$ channel is expected to be the dominant decay mode. The broad $B_{s2}(1D)$ and the narrow $B_{s2}^\prime(1D)$ correspond to the $2^-$ bottom-strange mesons belonging to the $(1^-,2^-)_{j=\frac{3}{2}}$ and $(1^-,2^-)_{j=\frac{5}{2}}$ doublets, respectively.

\begin{table}
\begin{center}
\caption{ \label{Bs1D} Decay widths of the $B_s(1D)$ states in MeV.}
\scriptsize
\begin{tabular}{lcccccc}
\hline\hline
                               & $B_s(1^3D_1)$      & $B_s(1^3D_3)$   & $B_{s2}(1D)$   & $B_{s2}^\prime(1D)$  \\\hline
  $B^+K^-$                     & 52.95              & 11.99           & $-$            & $-$      \\
  $B^0\bar{K}^0$               & 53.48              & 11.70           & $-$            & $-$   \\
  $B^{*+}K^-$                  & 29.36              & 11.04           & 90.12          & 38.53      \\
  $B^{*0}\bar{K}^0$            & 29.57              & 10.74           & 90.38          & 37.88        \\
  $B_s\eta$                    & 19.13              & 0.57            & $-$            & $-$    \\
  $B_s^*\eta$                  & 8.80               & 0.30            & 18.15          & 2.95   \\
  Total width                  & 213.38             & 46.33           & 198.64         & 79.37   \\
\hline\hline
\end{tabular}
\end{center}
\end{table}

\section{Radiative transitions}

Besides the strong decays, the radiative transitions are also the important tools to determine the properties of heavy-light mesons. We evaluate the $E1$  and $M1$ radiative partial widths between the $v=n^{2S+1}L_J$ and $v^\prime = n^{\prime 2S+1}L^\prime_{J^\prime}$ states using\cite{Close:2005se,Godfrey:2004ya,Godfrey:2005ww}
\begin{eqnarray}
\Gamma_{E1}(v \to v^\prime + \gamma) = \frac{4\alpha e^2_Q}{3}C_{fi}\delta_{SS^\prime}|<v^\prime|r|v>|^2 \frac{E^3_\gamma E_f}{M_i},
\end{eqnarray}
\begin{eqnarray}
&&\Gamma_{M1}(v \to v^\prime + \gamma) \nonumber\\
&&= \frac{\alpha e^{\prime 2}_Q}{3}\frac{2J^\prime +1}{2L+1}\delta_{LL^\prime}\delta_{SS^\prime \pm 1}|<v^\prime|j_0(\frac{E_\gamma r}{2})|v>|^2 \frac{E^3_\gamma E_f}{M_i},
\end{eqnarray}
where $e_Q=\frac{m_q Q_b +m_b Q_q}{m_q+m_b}$, $e^\prime _Q=\frac{m_q Q_b +m_b Q_q}{m_qm_b}$, $Q_b$ and $Q_q$ stand for the charges of the quark $b$ and $q$ in units of $|e|$, respectively. $\alpha = 1/137$ is the fine-structure constant, $E_\gamma$ is the photon energy, $E_f$ is the energy of final heavy-light meson, $M_i$ is the mass of initial state, and the angular matrix element $C_{fi}$ can be expressed as
\begin{eqnarray}
C_{fi} = \rm{Max}(L,L^\prime)(2J^\prime +1)
\left\{
\begin{array}{crrr}
L^\prime & J^\prime  & S\\
J & L & 1
\end{array}
\right\}^2.
\end{eqnarray}

The wavefunctions obtained from the quark model (\ref{Ham}) are used to calculate the $E1$  and $M1$ radiative partial widths. To determine the photon and final state energies, The masses of these initial and final states should be involved. For the well established $B$, $B^*$, $B_s$, and $B_s^*$, the masses are taken from PDG\cite{Agashe:2014kda}. For the $B_1^\prime(1P)$, $B(1^3P_2)$, $B(2^1S_0)$, $B(1^3D_3)$, $B_{s1}^\prime(1P)$, and $B_s(1^3P_2)$, their masses are taken to be the $B_1(5721)$, $B_2^*(5747)$, $B_J(5840)$, $B_J(5960)$, $B_{s1}(5830)$, and $B_{s2}^*(5840)$ masses, respectively. For other states, their masses are taken from the predictions of the quark model(\ref{Ham}). The $E1$ and $M1$ transitions widths of the neutral charge open-bottom states together with the photon energies are listed in Tables \ref{BE1},\ref{BM1},\ref{BsE1},\ref{BsM1}.

\begin{table}
\begin{center}
\caption{ \label{BE1} E1 transitions widths of neutral charge bottom mesons. $E_\gamma$ in MeV and $\Gamma$ in keV.}
\scriptsize
\begin{tabular}{lcccccc}
\hline\hline
  Multiplets                   & Initial meson      & Final meson         & $E_\gamma$     & $\Gamma$  \\\hline
  $B(2S)\to B(1P)$             & $B(2^3S_1)$        & $B(1^3P_0)$          & 250            & 21.4      \\
                               & $B(2^3S_1)$        & $B(1^3P_2)$         & 196            & 51.6   \\
                               & $B(2^3S_1)$        & $B_1(1P)$           & 206            & 11.67      \\
                               & $B(2^3S_1)$        & $B_1^\prime(1P)$    & 210            & 25.9        \\
                               & $B(2^1S_0)$        & $B_1(1P)$           & 176            & 49.6    \\
                               & $B(2^1S_0)$        & $B_1^\prime(1P)$    & 180            & 25.2   \\
  $B(1P)\to B(1S)$             & $B(1^3P_0)$        & $B(1^3S_1)$         & 347            & 116.9      \\
                               & $B(1^3P_2)$        & $B(1^3S_1)$         & 400            & 177.7   \\
                               & $B_1(1P)$          & $B(1^3S_1)$         & 390            & 53.1      \\
                               & $B_1^\prime(1P)$   & $B(1^3S_1)$         & 386            & 108.5       \\
                               & $B_1(1P)$          & $B(1^1S_0)$         & 432            & 130.2    \\
                               & $B_1^\prime(1P)$   & $B(1^1S_0)$         & 428            & 60.4   \\
  $B(1D)\to B(1P)$             & $B(1^3D_3)$        & $B(1^3P_2)$         & 248            & 127.0   \\
                               & $B(1^3D_1)$        & $B(1^3P_2)$         & 345            & 9.3   \\
                               & $B(1^3D_1)$        & $B(1^3P_0)$         & 398            & 283.5   \\
                               & $B(1^3D_1)$        & $B_1(1P)$           & 355            & 49.0   \\
                               & $B(1^3D_1)$        & $B_1^\prime(1P)$    & 359            & 106.2   \\
                               & $B_2(1D)$          & $B(1^3P_2)$         & 258            & 14.5   \\
                               & $B_2(1D)$          & $B_1(1P)$           & 269            & 143.1   \\
                               & $B_2(1D)$          & $B_1^\prime(1P)$    & 273            & 0.1   \\
                               & $B_2^\prime(1D)$   & $B(1^3P_2)$         & 362            & 57.2   \\
                               & $B_2^\prime(1D)$   & $B_1(1P)$           & 372            & 8.6   \\
                               & $B_2^\prime(1D)$   & $B_1^\prime(1P)$    & 376            & 356.3   \\
\hline\hline
\end{tabular}
\end{center}
\end{table}

\begin{table}
\begin{center}
\caption{ \label{BsE1} E1 transitions widths of the neutral charge bottom-strange mesons. $E_\gamma$ in MeV and $\Gamma$ in keV.}
\scriptsize
\begin{tabular}{lcccccc}
\hline\hline
  Multiplets                   & Initial meson        & Final meson           & $E_\gamma$     & $\Gamma$  \\\hline
  $B_s(2S)\to B_s(1P)$         & $B_s(2^3S_1)$        & $B_s(1^3P_0)$         & 242            & 17.2      \\
                               & $B_s(2^3S_1)$        & $B_s(1^3P_2)$         & 161            & 25.6   \\
                               & $B_s(2^3S_1)$        & $B_{s1}(1P)$          & 199            & 9.4      \\
                               & $B_s(2^3S_1)$        & $B_{s1}^\prime(1P)$   & 172            & 12.6        \\
                               & $B_s(2^1S_0)$        & $B_{s1}(1P)$          & 173            & 41.7    \\
                               & $B_s(2^1S_0)$        & $B_{s1}^\prime(1P)$   & 146            & 12.3   \\
  $B_s(1P)\to B_s(1S)$         & $B_s(1^3P_0)$        & $B_s(1^3S_1)$         & 330            & 84.7      \\
                               & $B_s(1^3P_2)$        & $B_s(1^3S_1)$         & 409            & 159   \\
                               & $B_{s1}(1P)$         & $B_s(1^3S_1)$         & 372            & 39.5      \\
                               & $B_{s1}^\prime(1P)$  & $B_s(1^3S_1)$         & 398            & 98.8        \\
                               & $B_{s1}(1P)$         & $B_s(1^1S_0)$         & 418            & 97.7    \\
                               & $B_{s1}^\prime(1P)$  & $B_s(1^1S_0)$         & 444            & 56.6   \\
  $B_s(1D)\to B_s(1P)$         & $B_s(1^3D_3)$        & $B_s(1^3P_2)$         & 251            & 113.2   \\
                               & $B_s(1^3D_1)$        & $B_s(1^3P_2)$         & 295            & 5.1   \\
                               & $B_s(1^3D_1)$        & $B_s(1^3P_0)$         & 374            & 204.4   \\
                               & $B_s(1^3D_1)$        & $B_{s1}(1P)$          & 332            & 35.3   \\
                               & $B_s(1^3D_1)$        & $B_{s1}^\prime(1P)$   & 305            & 56.8   \\
                               & $B_{s2}(1D)$         & $B_s(1^3P_2)$         & 242            & 10.5   \\
                               & $B_{s2}(1D)$         & $B_{s1}(1P)$          & 279            & 138.8   \\
                               & $B_{s2}(1D)$         & $B_{s1}^\prime(1P)$   & 253            & 0.04   \\
                               & $B_{s2}^\prime(1D)$  & $B_s(1^3P_2)$         & 311            & 31.5   \\
                               & $B_{s2}^\prime(1D)$  & $B_{s1}(1P)$          & 348            & 5.9   \\
                               & $B_{s2}^\prime(1D)$  & $B_{s1}^\prime(1P)$   & 321            & 195.4   \\
\hline\hline
\end{tabular}
\end{center}
\end{table}

\begin{table}
\begin{center}
\caption{ \label{BM1} M1 transitions widths of the neutral charge bottom mesons. $E_\gamma$ in MeV and $\Gamma$ in keV.}
\scriptsize
\begin{tabular}{lcccccc}
\hline\hline
  Initial Multiplet             & Initial meson      & Final meson          & $E_\gamma$     & $\Gamma$  \\\hline
  $B(1S)$                      & $B(1^3S_1)$        & $B(1^1S_0)$           & 45             & 0.1      \\
  $B(2S)$                      & $B(2^3S_1)$        & $B(1^1S_0)$           & 623            & 8.0   \\
                               & $B(2^3S_1)$        & $B(2^1S_0)$           & 31             & 0.05      \\
                               & $B(2^1S_0)$        & $B(1^3S_1)$           & 554            & 0.9        \\
  $B(1P)$                      & $B_1(1P)$          & $B(1^3P_0)$           & 46             & 0.03      \\
                               & $B_1^\prime(1P)$   & $B(1^3P_0)$           & 42             & 0.01   \\
                               & $B(1^3P_2)$        & $B_1(1P)$             & 11             & $1.4\times10^{-3}$      \\
                               & $B(1^3P_2)$        & $B_1^\prime(1P)$      & 15             & $1.7\times10^{-3}$      \\
  $B(1D)$                      & $B(1^3D_1)$        & $B_2(1D)$             & 90             & 0.7      \\
                               & $B_2^\prime(1D)$   & $B(1^3D_1)$           & 18             & $2.3\times10^{-3}$   \\
                               & $B_2^\prime(1D)$   & $B(1^3D_3)$           & 118            & 1.4   \\
                               & $B_2^\prime(1D)$   & $B_2(1D)$             & 108            & 1.9      \\
\hline\hline
\end{tabular}
\end{center}
\end{table}

\begin{table}
\begin{center}
\caption{ \label{BsM1} M1 transitions widths of the neutral charge bottom-strange mesons. $E_\gamma$ in MeV and $\Gamma$ in keV.}
\scriptsize
\begin{tabular}{lcccccc}
\hline\hline
  Initial Multiplet            & Initial meson        & Final meson             & $E_\gamma$     & $\Gamma$  \\\hline
  $B_s(1S)$                    & $B_s(1^3S_1)$        & $B_s(1^1S_0)$           & 48.9           & 0.1      \\
  $B_s(2S)$                    & $B_s(2^3S_1)$        & $B_s(1^1S_0)$           & 603            & 4.0   \\
                               & $B_s(2^3S_1)$        & $B_s(2^1S_0)$           & 25.9           & 0.02      \\
                               & $B_s(2^1S_0)$        & $B_s(1^3S_1)$           & 535            & 0.1        \\
  $B_s(1P)$                    & $B_{s1}(1P)$          & $B_s(1^3P_0)$          & 45             & 0.02      \\
                               & $B_{s1}^\prime(1P)$   & $B_s(1^3P_0)$          & 72             & 0.05   \\
                               & $B_s(1^3P_2)$        & $B_{s1}(1P)$            & 39             & 0.04      \\
                               & $B_s(1^3P_2)$        & $B_{s1}^\prime(1P)$     & 11             & $5.2\times10^{-4}$      \\
  $B_s(1D)$                    & $B_s(1^3D_1)$        & $B_{s2}(1D)$            & 55             & 0.1      \\
                               & $B_{s2}^\prime(1D)$   & $B_s(1^3D_1)$          & 17             & $1.3\times10^{-3}$   \\
                               & $B_{s2}^\prime(1D)$   & $B_s(1^3D_3)$          & 63             & 0.2   \\
                               & $B_{s2}^\prime(1D)$   & $B_{s2}(1D)$           & 72             & 0.4      \\
\hline\hline
\end{tabular}
\end{center}
\end{table}

From Tables \ref{BE1}, it can be seen that the $B(1^3P_0)\gamma$, $B_1(1P)\gamma$, and $B_1^\prime(1P)\gamma$ channels are essential to discriminate the $B(2^3S_1)$ and $B(1^3D_3)$ interpretations for the B(5970), since these these decay mode are forbidden for the $B(1^3D_3)$ state while allowable for the $B(1^3D_3)$ state. The $B(1^3P_2)\gamma$ final state for these two assignments has sizable decay widths, and can be observed experimentally.

\section{Summary}

In this paper, we calculate the bottom and bottom-strange meson spectroscopy in a nonrelativistic quark model proposed by Lakhina and Swanson. Our predictions, combined with the results from some other quark models, give the mass ranges of bottom and bottom-strange mesons. With these predictions, we give the possible quark-model assignments for these bottom mesons observed by LHCb and CDF Collaborations. Furthermore, the strong and radiative decay behaviors of these bottom and bottom-strange mesons are investigated with the realistic meson wave functions from our employed nonrelativistic quark model. The $B_1(5721)$ and $B^*_2(5747)$ can be classified into the $B_1^\prime(1P)$ and $B(1^3P_2)$, respectively. The $B_{s1}(5830)$ and $B^*_{s2}(5840)$ can be identified as the $B_{s1}^\prime(1P)$ and $B_s(1^3P_2)$ states, respectively. The $B_J(5840)$ and $B_J(5960)$ can be explained as the $B(2^1S_0)$ and $B(1^3D_3)$, respectively. The $B(5970)$ can be interpreted as the $B(2^3S_1)$ or $B(1^3D_3)$. The properties of other states are also predicted, which will be helpful to search for these states experimentally.

\bigskip
\noindent
\begin{center}
{\bf ACKNOWLEDGEMENTS}\\
\end{center}

We would like to thank Tim Gershon from the LHCb Collaboration and Yu-Bing Dong from IHEP for valuable discussions. This work is partly supported by the National Natural Science Foundation of China under Grant No. 11505158, the China Postdoctoral Science Foundation under Grant No. 2015M582197, the Postdoctoral Research Sponsorship in Henan Province under Grant No. 2015023,  and the Startup Research Fund of Zhengzhou University (Grants No. 1511317001 and No. 1511317002).

\end{document}